\documentclass[onecolumn,showpacs,floatfix,aps]{revtex4}
\usepackage{amssymb,amsmath}
\usepackage{subfigure}
\usepackage[dvips]{graphics,color}
\usepackage{epsfig}\usepackage{float}

\newcommand{\n}{\nonumber\\}

\newcommand{\bec}{\begin{center}}
\newcommand{\eec}{\end{center}}

\newcommand{\bea}{\begin{array}}
\newcommand{\ear}{\end{array}}

\newcommand{\bfr}{\begin{flushright}}

\newcommand{\efr}{\end{flushright}}
\newcommand{\noi}{\noindent}
\newcommand{\me}{\frac{1}{2}}

\newcommand{\cl}{{\mt{C}}\ell}

\newcommand{\RR}{\mathbb{R}}

\newcommand{\ot}{\otimes}
\newcommand{\la}{\Lambda}
\newcommand{\w}{\wedge}
\newcommand{\g}{\gamma}

\newcommand{\beq}{\begin{eqnarray}}\newcommand{\benu}{\begin{enumerate}}\newcommand{\enu}{\end{enumerate}}
\newcommand{\eeq}{\end{eqnarray}}
\newcommand{\mt}{\mathcal}

\newcommand{\pa}{\partial}

\newcommand{\CC}{\mathbb{C}}

\newcommand{\mk}{\mathfrak}

\newcommand{\bx}{\begin{pmatrix}}
\newcommand{\ex}{\end{pmatrix}}

\newcommand{\vt}{\vartheta}

\begin{document}

\title{The quadratic spinor Lagrangian, axial torsion current, and generalizations}
\author{R. da Rocha}
\email{roldao@ifi.unicamp.br}
\affiliation{Instituto de F\'{\i}sica Te\'orica, 
Universidade Estadual Paulista, 
Rua Pamplona 145\\
01405-900 S\~ao Paulo, Brazil\\and\\
DRCC - Instituto de F\'{\i}sica Gleb Wataghin\\ Universidade Estadual de Campinas
CP 6165, 13083-970 Campinas, Brazil}
\author{J. G. Pereira}
\email{jpereira@ift.unesp.br}
\affiliation{Instituto de F\'{\i}sica Te\'orica, 
Universidade Estadual Paulista, 
Rua Pamplona 145
01405-900 S\~ao Paulo, Brazil}


\begin{abstract}
We show that the Einstein-Hilbert, the Einstein-Palatini, and the Holst actions can be derived from the Quadratic Spinor Lagrangian (QSL), when the three classes of Dirac spinor fields, under Lounesto spinor field classification, are considered. To each one of these classes, there corresponds a unique kind of action for a covariant gravity theory. In other words, it is shown to exist a one-to-one correspondence between the three classes of \emph{non-equivalent} solutions of the Dirac equation, and Einstein-Hilbert, Einstein-Palatini, and Holst actions. Furthermore, it arises naturally, from  Lounesto spinor field classification, that any other class of spinor field --- Weyl, Majorana, flagpole, or flag-dipole spinor fields --- yields a trivial (zero) QSL, up to a boundary term. To investigate this boundary term we do not impose any constraint on the Dirac spinor field, and consequently we obtain new terms in the boundary component of the QSL. In the particular case of a teleparallel connection, an axial torsion 1-form current density is obtained. New terms are also obtained in the corresponding Hamiltonian formalism. We then discuss how these new terms could shed new light on more general investigations.
\end{abstract}
\maketitle

%%%%%%%%%%%%
\section{Introduction}

Spinor fields can be classified according to the values assumed by their respective bilinear covariants. There are only six classes of spinor fields \cite{lou1,lou2,holl}: three of them are related to the three non-equivalent classes of Dirac spinor fields (DSFs), and the others are constituted respectively by the so-called 
flag-dipole, flagpole and Weyl spinor fields \cite{lou1,lou2,holl}. 
Majorana and ELKO (\emph{Eigenspinoren des Ladungskonjugationsoperators}, or dual-helicity eigenspinors of the charge conjugation operator) spinor fields are special subclasses of flagpole spinor fields \cite{ro1}. By using one specific class of DSF, and imposing a condition of constant spinor field, it has already been shown that the Einstein-Hilbert Lagrangian of general relativity (GR), as well as the Lagrangian of its teleparallel equivalent (GR$_\|$), can be recast as a quadratic spinor Lagrangian (QSL) \cite{tn1,tung1,tung3}. This development was purposed in a tentative to better understand the question of the gravitational energy-momentum localization. Also, the QSL which is originally constructed with a $\mathfrak{sl}(2,\mathbb{C})$-valued connection, was generalized in \cite{jaco} to a more general 
$\mathfrak{gl}(2,\mathbb{C})$-valued connection.

In order to prove the
equivalence between the QSL and the Lagrangians associated with GR and GR$_\|$, a DSF of class-(2) --- under Lounesto spinor field classification --- with constant coefficients was used \cite{tn1}. Although this is a natural choice in the context of the QSL formalism of gravitational theory, it remains to be better justified. Furthermore, the use of spinor fields with constant coefficients is quite restrictive. It is true that one can force the DSF to have constant coefficients. This is possible because both the orthonormal frame field and the DSF symmetries, under Lorentz transformations, can be tied together \cite{tn1}. However, there are many other possible choices that \emph{do not} require the orthonormal frame gauge freedom to be the same as the DSF gauge freedom. In these cases, the rules of the Clifford algebra-valued differential forms imply the existence of extra terms in  the boundary term associated with the QSL. 
 
One of the main purposes of this paper is to establish an equivalence between the underlying algebraic structure of the DSFs and the corresponding gravity theory actions. This equivalence enables us to better characterize and understand the nature of the spinor field that constitutes the QSL. We show that, when the DSF is not restricted to the case of a class-(2) DSF with constant coefficients, in addition to the Einstein-Hilbert, also the Einstein-Palatini and the Holst actions can be derived from a QSL, when we consider class-(1) and -(3) DSFs.
We begin by showing first that the spinor-valued 1-form field entering the QSL has necessarily to be constructed by a tensor product between a \emph{Dirac spinor field} and a Clifford algebra-valued 1-form: no other spinor fields can lead either to the Holst action, or to the particular cases of Einstein-Hilbert and Einstein-Palatini actions. These three gravitational actions correspond respectively to a class-(2), class-(3), and class-(1) DSFs.  Classes-(2) and -(3) together give the  Einstein-Hilbert and Einstein-Palatini actions, and class-(1) DSF gives alone the complete Holst action. We mention in passing that this action shows up also in the proof of gravitational theory as a SUSY gauge theory \cite{tung2}. 
Furthermore, we assume a more general approach, where the DSF is not a constant spinor field anymore. As a consequence, the boundary term of the QSL will have many additional terms that can be related to some physical identities, and may unravel additional properties.

The paper is organized as follows: after presenting some algebraic preliminaries in Section \ref{w2}, we investigate in Section \ref{w3} the extra terms in the boundary term of the QSL. In the particular case of a teleparallel spacetime, these extra terms give rise to an axial torsion current density, coupling the 1-form axial torsion and the DSF through its total derivative $d\psi$. In Section \ref{ui}, after briefly presenting the Lounesto spinor field classification, as well as some important features of each spinor field class, we show that Einstein-Hilbert, Einstein-Palatini, and Holst actions can be derived from a QSL provided we do not restrict ourselves to the case of a class-(2) DSF. In Section \ref{w5} we obtain the extra terms in the corresponding Hamiltonian formalism. Some discussions concerning these generalizations are presented in the last Section. 
 
%%%%%%%%%%%%
\section{Preliminaries}
\label{w2}
We denote by $\mathcal{M=} (M, g,\nabla,\tau_g,\uparrow)$ the spacetime structure: $M$ denotes a 4-dimensional manifold, $g
\in\sec T_{0}^{2}M$ is the metric associated with the cotangent bundle, $\nabla$ is the Levi-Civita connection
of $
g$, $\tau_g\in\sec
{\displaystyle\Lambda^{4}}
(T^{\ast}M)$ defines a spacetime orientation and $\uparrow$ refers to an equivalence class of timelike 1-form fields defining a time orientation.  By
$F(M)$ we mean the (principal) bundle of frames, by $\mathbf{P}%
_{\mathrm{SO}_{1,3}^{e}}(M\mathbf{)}$ the orthonormal frame bundle, and
$P_{\mathrm{SO}_{1,3}^{e}}(M)$ denotes the orthonormal coframe bundle. We consider $M$ a spin manifold, and then there exists $\mathbf{P}_{\mathrm{Spin}%
_{1,3}^{e}}(M\mathbf{)}$ and $P_{\mathrm{Spin}_{1,3}^{e}}(M\mathbf{)}$ which
are respectively the spin frame and the spin coframe bundles. We denote by
$s:P_{\mathrm{Spin}_{1,3}^{e}}(M\mathbf{)\rightarrow}P_{\mathrm{SO}_{1,3}^{e}%
}(M\mathbf{)}$ the fundamental mapping present in the definition of
$P_{\mathrm{Spin}_{1,3}^{e}}(M\mathbf{)}$. 
A spin structure on $M$ consists of a principal fiber
bundle $\mathbf{\pi}_{s}:P_{\mathrm{Spin}_{1,3}^{e}}(M)\rightarrow M$, 
with group $\mathrm{Spin}_{1,3}^{e}$, and the map
\begin{equation}
s:P_{\mathrm{Spin}_{1,3}^{e}}(M)\rightarrow P_{\mathrm{SO}_{1,3}^{e}%
}(M)\label{spinor bundle 1}%
\end{equation}
satisfying the following conditions:

(i) $\mathbf{\pi}(s(p))=\mathbf{\pi}_{s}(p),\ \forall p\in P_{\mathrm{Spin}%
_{1,3}^{e}}(M);$ $\pi$ is the projection map of the bundle $P_{\mathrm{SO}%
_{1,3}^{e}}(M)$.

(ii) $s(p \phi)=s(p)\mathrm{Ad}_{\phi},\;\forall p\in P_{\mathrm{Spin}_{1,3}^{e}}(M)$ and
$\mathrm{Ad}:\mathrm{Spin}_{1,3}^{e}\rightarrow\mathrm{Aut}(\cl_{1,3}),$
$\mathrm{Ad}_{\phi}:\cl_{1,3}\ni \Xi\mapsto \phi\Xi\phi^{-1}\in\cl_{1,3}$ \cite{moro}.

We recall now that sections of $P_{\mathrm{SO}_{1,3}^{e}%
}(M\mathbf{)}$ are orthonormal coframes, and that sections of $P_{\mathrm{Spin}%
_{1,3}^{e}}(M\mathbf{)}$ are also orthonormal coframes such that two coframes
differing by a $2\pi$ rotation are distinct and two coframes differing by a
$4\pi$ rotation are identified. 
Next we introduce the Clifford bundle of differential forms $\mathcal{C\ell
(}M,g)$, which is a vector bundle associated with $P_{\mathrm{Spin}%
_{1,3}^{e}}(M\mathbf{)}$. Their sections are sums of non-homogeneous
differential forms, which will be called Clifford fields. We recall that
\ $\mathcal{C\ell(}M,g)=P_{\mathrm{SO}_{1,3}^{e}}(M)\times
_{\mathrm{Ad}^{\prime}}\cl_{1,3}$, where $\cl_{1,3}%
\simeq$ M(2,${\mathbb{H}})$ is the spacetime algebra. Details of the bundle
structure are as follows \cite{dimakis,dimakis1,est}:

(1) Let $\mathbf{\pi}_{c}:\mathcal{C}\ell(M,g)\rightarrow M$ be
the canonical projection of $\mathcal{C}\ell(M,g)$ and let
$\{U_{\alpha}\}$ be an open covering of $M$. There are trivialization mappings
$\mathbf{\psi}_{i}:\mathbf{\pi}_{c}^{-1}(U_{i})\rightarrow U_{i}%
\times\cl_{1,3}$ of the form $\mathbf{\psi}_{i}(p)=(\mathbf{\pi}%
_{c}(p),\psi_{i,x}(p))=(x,\psi_{i,x}(p))$. If $x\in U_{i}\cap U_{j}$ and
$p\in\mathbf{\pi}_{c}^{-1}(x)$, then
\begin{equation}
\psi_{i,x}(p)=h_{ij}(x)\psi_{j,x}(p)
\end{equation}
for $h_{ij}(x)\in\mathrm{Aut}(\cl_{1,3})$, where $h_{ij}:U_{i}\cap
U_{j}\rightarrow\mathrm{Aut}(\cl_{1,3})$ are the transition mappings of
$\mathcal{C}\ell(M,g)$. We recall that every automorphism of
$\cl_{1,3}$ is \textit{inner. }Then,
\begin{equation}
h_{ij}(x)\psi_{j,x}(p)=a_{ij}(x)\psi_{i,x}(p)a_{ij}(x)^{-1} \label{4.4}%
\end{equation}
for some $a_{ij}(x)\in\cl_{1,3}^{\star}$, the group of invertible
elements of $\cl_{1,3}$.

(2) As it is well known, the group $\mathrm{SO}_{1,3}^{e}$ has a natural
extension in the Clifford algebra $\cl_{1,3}$. Indeed, we know that
$\cl_{1,3}^{\star}$ (the group of invertible elements of $\cl%
_{1,3}$) acts naturally on $\cl_{1,3}$ as an algebra automorphism
through its adjoint representation. A set of \emph{lifts} of the transition
functions of $\mathcal{C}\ell(M,\mathtt{g})$ is a set of elements
$\{a_{ij}\}\subset$ $\cl_{1,3}^{\star}$ such that, if \footnote{Recall
that $\mathrm{Spin}_{1,3}^{e}=\{\phi\in\cl_{1,3}^{0}:\phi\tilde{\phi}%
=1\}\simeq\mathrm{SL}(2,\mathbb{C)}$ is the universal covering group of the
restricted Lorentz group $\mathrm{SO}_{1,3}^{e}$. Notice that $\cl%
_{1,3}^{0}\simeq\cl_{3,0}\simeq$ M(2,$\mathbb{C})$, the even subalgebra of
$\cl_{1,3}$ is the Pauli algebra.}
\begin{eqnarray}
&&\mathrm{Ad} :\phi\mapsto\mathrm{Ad}_{\phi} \nonumber \\
&&\mathrm{Ad}_{\phi}(\Xi) =\phi \Xi\phi^{-1}, \quad \forall \Xi\in\cl_{1,3},
\end{eqnarray}
then $\mathrm{Ad}_{a_{ij}}=h_{ij}$ in all intersections.

(3) Also $\sigma=\mathrm{Ad}|_{\mathrm{Spin}_{1,3}^{e}}$ defines a group
homeomorphism $\sigma:\mathrm{Spin}_{1,3}^{e}\rightarrow\mathrm{SO}_{1,3}^{e}$
which is onto with kernel $\mathbb{Z}_{2}$. We have that Ad$_{-1}=$ identity,
and so $\mathrm{Ad}:\mathrm{Spin}_{1,3}^{e}\rightarrow\mathrm{Aut}%
(\cl_{1,3})$ descends to a representation of $\mathrm{SO}_{1,3}^{e}$.
Let us call $\mathrm{Ad}^{\prime}$ this representation, i.e., $\mathrm{Ad}%
^{\prime}:\mathrm{SO}_{1,3}^{e}\rightarrow\mathrm{Aut}(\cl_{1,3})$.
Then we can write $\mathrm{Ad}_{\sigma(\phi)}^{\prime}\Xi=\mathrm{Ad}_{\phi}%
\Xi=\phi\Xi\phi^{-1}$.

(4) It is clear that the structure group of the Clifford bundle
$\mathcal{C}\ell(M,g)$ is reducible from $\mathrm{Aut}%
(\cl_{1,3})$ to $\mathrm{SO}_{1,3}^{e}$. The transition maps of
the principal bundle of oriented Lorentz cotetrads $P_{\mathrm{SO}_{1,3}^{e}%
}(M)$ can thus be (through $\mathrm{Ad}^{\prime}$) taken as transition maps for the
Clifford bundle. We then have \cite{lawmi}
\begin{equation}
\mathcal{C}\ell(M,g)=P_{\mathrm{SO}_{1,3}^{e}}(M)\times
_{\mathrm{Ad}^{\prime}}\cl_{1,3},
\end{equation}
i.e., the Clifford bundle is a vector bundle associated with the principal
bundle $P_{\mathrm{SO}_{1,3}^{e}}(M)$ of orthonormal Lorentz coframes.

Recall that $\mathcal{C}\!\ell(T_{x}^{\ast}M,g_{x})$ is also a
vector space over $\mathbb{R}$ which is isomorphic to the exterior algebra
$\Lambda (T_{x}^{\ast}M)$ of the cotangent space and $\Lambda (T_{x}^{\ast}M)=%
%TCIMACRO{\dbigoplus \nolimits_{k=0}^{4}}%
%BeginExpansion
{\displaystyle\oplus_{k=0}^{4}}
%EndExpansion
\Lambda{}^{k}(T_{x}^{\ast}M)$, where $\Lambda^{k}(T_{x}^{\ast}M)$ is the
$\binom{4}{k}$-dimensional space of $k$-forms over a point $x$ on $M$. There is a natural embedding
\ $\Lambda (T^{\ast}M)\hookrightarrow$ $\mathcal{C}\ell(M,g)$
\cite{lawmi} and sections of $\mathcal{C}\!\ell(M,g)$ --- Clifford
fields --- can be represented as a sum of non-homogeneous differential forms.
Let $\{e_{a}\}\in\sec\mathbf{P}_{\mathrm{SO}_{1,3}^{e}}(M)$
(the orthonormal frame bundle) be a tetrad basis for $TU\subset TM$ (given an open set $U\subset M$). Moreover, let
$\{\vartheta^{a}\}\in\sec P_{\mathrm{SO}_{1,3}^{e}}(M)$.
Then, for each ${a}=0,1,2,3$, ${\vartheta}^{a}%
\in\sec\Lambda^{1}(T^{\ast}M)\hookrightarrow\sec\mathcal{C}\!\ell
(M,g$).
We recall next the crucial result \cite{moro,lawmi} that in a spin manifold we
have:
\begin{equation}
\mathcal{C}\ell(M,\mathtt{\eta})=P_{\mathrm{Spin}_{1,3}^{e}}(M)\times
_{\mathrm{Ad}}\cl_{1,3}. \label{1new}%
\end{equation}
Spinor fields are sections of vector bundles associated with  the principal
bundle of spinor coframes. The well known Dirac spinor fields are sections of
the bundle
\begin{equation}
S_{c}(M,\mathtt{\eta})=P_{\mathrm{Spin}_{1,3}^{e}}(M)\times_{\mu_{c}}\mathbb{C}^{4}, \label{4.7}%
\end{equation}
with $\mu_{c}$ the $D^{(1/2,0)}\oplus D^{(0,1/2)}$ representation of $\mathrm{Spin}%
_{1,3}^{e}\cong\mathrm{SL}(2,\mathbb{C})$ in $\mathrm{End}(\mathbb{C}^{4})$
~\cite{choquet}.

 %Let $M$ be a 4-dimensional manifold endowed with a Lorentzian metric $g:\sec\Lambda^1(T^*M)\times\sec\Lambda^1(T^*M)
%\rightarrow \RR$
%of index two and 
% [$\{e_a\}\in\sec\la^1(T^*M)$]
The orthonormal coframe field $\{\vartheta^a\}\in\sec\la^1(T^*M)$ can be related  
to the metric $g$ by $g=\eta_{ab}\vartheta^a\ot\vartheta^b$, 
with $(\eta_{ab}) = {\rm diag} (1,-1,-1,-1)$. In other words, $g$ is the metric on $M$ according to
which the elements of $\{e_{a}\}$ are orthonormal vector fields,
i.e., $g_{x}(e_{a}|_{x},e_{b}|_{x}) = \eta_{ab}$ for each $x \in
M$. We use the Latin alphabet $a, b, c, \ldots = 0, 1,
2, 3$ to denote anholonomic indices related to the tangent
spaces and the Greek
alphabet $\mu, \nu, \rho, \ldots = 0, 1, 2, 3$ to denote holonomic
spacetime indices. Let
$\{x^{\mu}\}$ be local coordinates in an open set $U\subset M$. Denoting
$\partial_{\mu}=\partial/\partial x^{\mu}$, one can always expand
the coordinate basis $\{\partial_{\mu}\}$ in terms of $\{e_{a}\}$,
\[
\partial_{\mu} = h^{a}{}_{\mu} e_{a}
\]
for certain functions $h^{a}{}_{\mu}$ on $U$. This immediately yields
$g_{\mu\nu} := g(\partial_{\mu},\partial_{\nu}) = h^{a}{}_{\mu}h^{b}{}_{\nu}
\eta_{ab}$. 
Consider a Minkowski vector space $V = \RR^{1,3}$, 
isomorphic (as a vector space) to $T_xM$ and its associated Clifford algebra $\cl_{1,3}$, 
generated by the basis $\{\gamma_\mu\}$ and by the relations $\gamma_\mu\gamma_\nu + \gamma_\nu\gamma_\mu = 
2\eta_{\mu\nu}$. The Clifford product will be denoted by juxtaposition. The pseudoscalar $\gamma_5$ is defined as $\g_5=\g_{0123}$.
Given two arbitrary (in general non-homogeneous) form fields $\xi, \zeta\in\sec\Lambda(T^*M)$, 
the dual Hodge operator $\star:\sec\Lambda^p(T^*M)
\rightarrow\sec\Lambda^{4-p}(T^*M)$ is defined explicitly by 
$\xi\w\star\zeta = G(\xi,\eta)$, where $G: \sec\Lambda(T^*M)\times\sec\Lambda(T^*M)
\rightarrow \RR$ denotes the metric extended to the space of form fields.

The coframe field $\{\vartheta^a\}$ and the metric-compatible connection 1-form field $\omega^{ab}$ 
are potentials for the curvature and the torsion, expressed respectively by 
the structure equations
\begin{equation}\label{2}
\Omega^a_{\;\;b}=d \omega^a_{\;\;b} + 
\omega^a_{\;\;\rho}\w \omega^\rho_{\;\;b}\in\sec\la^2(T^*M) 
\quad \mbox{and} \quad \Theta^a = d\vartheta^a + \omega^a_{\;\;b}\wedge \vartheta^b\in\sec\la^2(T^*M).
\end{equation}
The connection 
coefficients are implicitly given by $\omega_{ab}=\omega_{abc}\theta^c$, and the torsion can be decomposed in its irreducible components under the global Lorentz group as \cite{h2}
\begin{equation}
\Theta^a = {}^{(1)}\Theta^a + {}^{(2)}\Theta^a + {}^{(3)}\Theta^a
\end{equation}
where 
\begin{equation}\label{at3}
{}^{(2)}\Theta^a = \frac{1}{3}\vartheta^a\w(\vartheta^b\lrcorner \Theta_b),
\quad  {}^{(3)}\Theta^a = -\frac{1}{3}\star(\vartheta^a\w \mk{a}),\quad  
 {}^{(1)}\Theta^a = \Theta^a -  {}^{(2)}\Theta^a  - {}^{(3)}\Theta^a,
\end{equation}
with $\mk{a} = \star(\Theta_b\w\vartheta^b)$ denoting the axial 1-form associated with the axial torsion ${}^{(3)}\Theta^a$. 
The term $\star{\mk{a}}$ is the well known translational Chern-Simons 3-form field \cite{h1,h2,h3}, 
whose total derivative $d\star\mk{a}$ is the Nieh-Yan 4-form field \cite{z1,z2,ni}. 

Clifford algebra-valued differential forms (on Minkowski spacetime) 
are elements of $\sec \la(T^*M)\ot\cl_{1,3}$. In particular,
Eqs.~(\ref{2}) are written as
\begin{equation}\label{4}
\Omega = d\omega + \omega\w\omega\ \quad \mbox{and} \quad
\Theta = d\vartheta + \omega\w\vartheta + \vartheta\w\omega,
\end{equation}
where
\beq\label{3}
\vartheta &=& \vartheta^a\ot\g_a,\qquad \omega = \frac{1}{4}\omega^{ab}\ot\gamma_{ab},\n
\Theta &=& \Theta^a\ot\g_a,\qquad \Omega = \frac{1}{4}\Omega^{ab}\ot\gamma_{ab},
\eeq
with $\gamma_{ab}=\me(\gamma_a\g_b - \g_b\g_a)$.
All operations in the exterior algebra of differential forms 
 are trivially induced on the space of Clifford-valued differential forms.
In particular, given $\phi^a\in\Lambda(V)$,  the total derivative $d(\phi^a\ot\g_a)$ is given by $d(\phi^a)\ot\g_a$ and, given
a $p$-form field basis $\{\vt^I\}$ and a Clifford algebra basis 
$\{\g_I = \g_a\g_b\g_c\ldots\}$, the exterior product between two elements
$\Phi = \Phi^I\ot\g_I$ and $\Gamma = \Gamma^J\ot\g_J$ of $\sec\la(T^*M)\ot\cl_{1,3}$ is given by \cite{dimakis,est}
\begin{equation}
\Phi\w\Gamma = (\Phi^I\ot\g_I)\w(\Gamma^J\ot\g_J) = (\Phi^I\w\Gamma^J)\ot\g_I\g_J.
\end{equation}

%%%%%%%%%%%%%%%%%%%%%%%
\section{The quadratic spinor Lagrangian}
\label{w3}

Given a spinor-valued 1-form field $\Psi$, the quadratic spinor Lagrangian (QSL) is given by 
\begin{equation}\label{13}
\mathcal{L}_\Psi = 2D\bar{\Psi}\w\g_5 D\Psi = 2\bar\Psi\w\Omega\g_5\w\Psi
 + d[(D\bar\Psi)\w\g_5\Psi + \bar\Psi\w \g_5
D\Psi],
\end{equation}\noi
where
\begin{equation}\label{00}
D\Psi = d\Psi + \omega\w\Psi\qquad {\rm and} \qquad D\bar\Psi = d\bar\Psi + \bar\Psi\w\omega. 
\end{equation}\noi 
Now, choose the \emph{ansatz} 
\begin{equation}\label{35}\Psi = \psi\ot\vartheta,\end{equation} 
where $\vartheta$ denotes the orthonormal frame 1-form $\vartheta = \vartheta^a\ot\g_a = h^a_{\;\,\mu} dx^\mu\ot\g_a$ and $\psi$ is a spinor field. The action of the spinor covariant exterior derivative $D$, mapping a spinor-valued 1-form field $\Psi$ 
into a spinor-valued 2-form field $D\Psi$ is explicitly given by\footnote{More generally, the spinor 
covariant exterior derivative $D$  maps a spinor-valued $p$-form field  
to a spinor-valued $(p+1)$-form field.}
$$D\Psi = \vartheta^a\w[\pa^{(s)}\psi\ot\vartheta + \psi\ot(\nabla_{e_a} + (e_a\lrcorner \Theta^c)\w e_c\lrcorner) \vartheta],
$$
where the spin-Dirac operator $\pa^{(s)}$ acting on spinor fields $\psi$ and the covariant derivative $\nabla_{e_a}$
acting on Clifford-valued 1-form fields are given respectively by
\beq
\pa^{(s)}\psi &=& \pa_a\psi + \frac{1}{2}\omega_a\psi,\n
\nabla_{e_a}\vartheta &=& \pa_a\vartheta + \frac{1}{2}[\omega_a,\vartheta],
\eeq
where $\omega_a = \omega_{ac}^{b}(e_b\ot\vartheta^c)$.

The \emph{ansatz} given by Eq.(\ref{35}) arises 
in different contexts: in \cite{abelow} $\psi$ is a  Dirac spinor field
used to prove the equivalence between QSL and  the Lagrangians describing General Relativity (GR), 
its teleparallel equivalent GR$_\|$, and the M\o ller Lagrangian; in \cite{tung2} $\psi$ is an auxiliary Majorana spinor used to prove 
that gravitation can be described as a SUSY gauge theory; in \cite{bars,bars1} $\psi$ is an \emph{anticommuting} Majorana spinor described by Grassmann superspace coordinates, which generates the spinor supersymmetric conserved current. 
The QSL was first proposed in \cite{tn1} in the proof of positive energy theorem.

Up to our knowledge, there are no identities like the spinor-curvature identities that yield the term linear 
in curvature which reduces to the scalar curvature \cite{tung3}.  One of the spinor-curvature identities is given by
\begin{equation}
2D(\bar\psi \xi)\w D(\zeta\psi) = 2(-1)^p\bar\psi \xi\w\Omega\w(\zeta\psi) + d[\bar\psi\xi\w D(\zeta\psi) - (-1)^p D(\bar\psi\xi)\w \zeta\psi],
\end{equation}
where now $\xi\in\sec\Lambda^p(T^*M)\ot\cl_{1,3}$ and $\zeta\in\sec\Lambda(T^*M)\ot\cl_{1,3}$.
The scalar curvature appears in a natural way in the case where $\Psi$ in QSL is a spinor-valued 1-form field, like in Eq.(\ref{35}), as we shall see below.

Substituting  the \emph{ansatz} (\ref{35}) in the QSL  (Eq.(\ref{13})), it follows
\beq
\mathcal{L}_\Psi = 
\mathcal{L}(\psi,\vartheta,\omega) &=& 2D(\bar{\psi}\vartheta)\g_5\w D(\vartheta\psi) \nonumber\\
&=&-\bar\psi\psi\,\Omega_{ab}\w\star(\vartheta^a\w\vartheta^b) + 
\bar\psi\g_5\psi\,\Omega_{ab}\w\vartheta^a\w\vartheta^b + d[D(\bar\psi\vartheta)\g_5\psi\vartheta + \bar\psi\vartheta \g_5 D(\vartheta\psi)].
\eeq\noi When the spinor field satisfies the normalization conditions 
\begin{equation}\label{5}
\bar\psi\psi = 1, \qquad\qquad\bar\psi\g_5\psi = 0,\end{equation}\noi
the original QSL can be written as 
\begin{equation}\label{6}
\mathcal{L}_\Psi = 
-\Omega_{ab}\w\star(\vartheta^a\w\vartheta^b) + 
d[D(\bar\psi\vartheta)\w\g_5\psi\vartheta + \bar\psi\vartheta \w\g_5  D(\vartheta\psi)]
\end{equation}\noi
The DSF $\psi$ enters in the QSL only at the boundary and does not appear in the equations of motion. 
Up to the boundary term, the Lagrangian  is given by 
\begin{equation}
\mt{L}_\Psi = -\Omega_{ab}\w\star(\vartheta^a\w\vartheta^b),
\end{equation}\noi which is the Einstein-Hilbert Lagrangian.
 Eq.(\ref{6}) shows that the action $S_\Psi = \int\mt{L}_\Psi$ does not depend locally on the Dirac spinor field $\psi$.  

Tung and Nester \cite{abelow} asserted that a change on the spinor field will leave the action $S_\Psi$
	 unchanged, and
then the spinor field has a six-parameter ---
 four complex parameters constrained by Eqs.(\ref{5}) --- local gauge invariance. The
theory also presents a Lorentz gauge freedom related to the transformations of the orthonormal frame field. They
prove that the spinor field gauge freedom induces a Lorentz transformation on the orthonormal frame field, and the
boundary term has only one physically independent degree of freedom \cite{tn1}. They also admit a suitable choice fixing one
of the two Lorentz gauges by tying the DSF to the orthonormal coframe field together. So, the spinor
gauge freedom related to the six parameter DSF $\psi$ is (2 to 1)
 equivalent to the Lorentz transformations for the
associated orthonormal frame. The choice $d\psi = 0$ clearly implies that $\psi$
 is a constant spinor. However, other
choices are possible where the spinor field $\psi$ is not constant anymore --- $d\psi\neq 0$. There are cases (see Section (IV))
where the spinor field gauge freedom is entirely independent on the Lorentz transformations of the orthonormal frames. In the most general case, when $d\psi\neq 0$, in addition to the well known boundary terms \cite{tn1} 
\begin{equation}
-\bar\psi\psi\,\omega_{ab}\w\star(\vartheta^a\w\vartheta^b) + \bar\psi\g_5\psi\,\omega_{ab}\w\vartheta^a\w\vartheta^b
+ \bar\psi\g_5\psi\, d\vartheta^a\w e_a,
\end{equation}\noi
there are extra terms given by (differentiating the relation $\bar\psi\g_5\psi = 0$
to obtain $d\bar\psi\g_5\psi = -\bar\psi\g_5 d\psi$)
\begin{equation}
\mbox{``extra terms''} = d\bar\psi\g_5(d\vartheta^a\w\vartheta^b\w\omega_{ab} +  d\omega_{ab}\w \vartheta^a\w
\vartheta^b)\psi + d\bar\psi\g_5(\omega_{ab}\w\vartheta^a\w\vartheta^b - d\vartheta^a\w \vartheta_a)d\psi. 
\end{equation}
In general, the extra terms above do not have a clear interpretation. In the teleparallel case, however, where $\omega = 0$, the extra terms reduce to 
 \begin{equation}\label{14}
- \bar\varphi\g_5(\star\mk{a})\varphi,
\end{equation}\noi
where $\varphi:= d\psi$, and $\mk{a} = \mk{a}^b\ot\g_b = -\star(\vartheta^a\w \Theta_a)$ denotes the axial 1-form field associated with the 
axial torsion 2-form ${}^{(3)}\Theta^a$, given by Eq. (\ref{at3}). 
In the teleparallel case it is described by the frame 
anholonomy, since in this case $\Theta^a = d\vartheta^a$. The axial 1-form field is the so-called
translational Chern-Simons term, whose total derivative is the Nieh-Yan 4-form field \cite{z1,ni}.

The term in Eq.(\ref{14}) suggests a coupling between the axial 1-form field $\mk{a}$ and the Dirac spinor $\psi$
field through its total derivative $d\psi$. In other words, if we suppose that $d\psi\neq 0$, there appears on the
boundary an extra axial torsion 1-form field current density
\begin{equation}
\star\mk{a}^b\bar\varphi\g_5\g_{b}\varphi.
\end{equation}
In the next Section we will show how to obtain the Holst action, which consists in the sum of Einstein-
Hilbert and Einstein-Palatini actions. This will be done by relaxing the normalization conditions (\ref{5}). We also \emph{prove} that the spinor field in Eq. (\ref{35}) must be a DSF in order to make the QSL to yield the Einstein-Hilbert, Einstein-Palatini and Holst actions. Any other choice of spinor field --- like Majorana, Weyl, flag-dipole, flagpole, pure spinor fields --- up to a boundary term, gives a trivial QSL.

%%%%%%%%%%%%%%%%%%%%%%%%%%%%%%%%%%%%%%%%%%%%%%%%%%

\section{QSL as the fundament of gravity via the classification of spinors}
\label{ui}
Classical spinor fields\footnote{As is well known, quantum spinor fields are operator valued distributions. It is not necessary to introduce quantum fields in order to
know the algebraic classification of ELKO spinor fields.} carrying a $D(1/2,0)\oplus D(0,1/2)$, or $D(1/2,0)$, or $D(0,1/2)$ representation of SL$(2,\CC) \simeq$ Spin$^e_{1,3}$ are sections of the vector bundle
\begin{equation}
P_{{\rm Spin}^e_{1,3}} (M) \times_\rho \CC^4,\end{equation}
\noi where $\rho$ stands for the $D(1/2,0) \oplus D(0,1/2)$ (or $D(1/2,0)$ or $D(0,1/2)$) representation of Spin$^e_{1,3}$ in $\CC^4$. Other important spinor fields, like Weyl spinor fields, are obtained by imposing some constraints on the sections
of $P_{{\rm Spin}^e_{1,3}}(M) \times_\rho \CC^4$. See, e.g., \cite{lou1,lou2} for details.
Given a spinor field $\psi$ $\in
\sec\mathbf{P}_{\rm{Spin}_{1,3}^{e}}(M)\times_{\rho}\mathbb{C}^{4}$ the
bilinear covariants are the following sections of ${\displaystyle\Lambda
}(T^*M)={\displaystyle\oplus_{p=0}^{4}}$ ${\displaystyle\Lambda^{p}%
}(T^*M)\hookrightarrow C\mathcal{\ell}(M,g)$ \cite{ro1,moro}
\begin{align}
\sigma &  =\bar\psi\psi,\quad\mathbf{J}=J_{\mu}\vartheta%
^{\mu}=\bar\psi\gamma_{\mu}\psi\,\vartheta^{\mu},\quad
\mathbf{S}=S_{\mu\nu}\vartheta^{\mu\nu}=\frac{1}{2}\psi^{\dagger}\gamma
_{0}i\gamma_{\mu\nu}\psi\,\vartheta^{\mu}\wedge\vartheta^{\nu},\nonumber\\
\mathbf{K} &  =\bar\psi i\gamma_{0123}\gamma_{\mu}
\psi\,\vartheta^{\mu},\quad\chi=-\bar\psi\gamma_{0123}%
\psi,\label{fierz}%
\end{align}
with $\sigma,\chi\in\sec
\Lambda^0
(T^*M),\, \mathbf{J,K}\in\sec\Lambda^1
(T^*M)$ and $\mathbf{S}\in\sec
\Lambda^2(T^*M)\hookrightarrow C\mathcal{\ell}(M,g)$. In the formul\ae\, appearing in
Eq. (\ref{fierz}), the set $\{\gamma_{\mu}\}$ can be thought of as being the Dirac matrices, but we prefer not to make reference to any
kind of representation, in order to preserve the algebraic character of the theory. When required, it is possible
to use any suitable representation. Also, $
\{\mathbf{1}_{4},\gamma_{\mu},\gamma_{\mu}\gamma_{\nu},\gamma_{\mu}\gamma
_{\nu}\gamma_{\rho},\gamma_{0}\gamma_{1}\gamma_{2}\gamma_{3}\}$ 
is a basis for  $C\mathcal{\ell}(M,g)$, $\mu<\nu<\rho$, and
 $\mathbf{1}_{4}\in$ $\mathbb{C(}4\mathbb{)}$ is the identity matrix.

In the case of the electron, described by DSFs
(classes (1), (2) and (3) below), $\mathbf{J}$ is a future-oriented timelike current
1-form field which gives the current of probability, the 2-form field $\mathbf{S}$ is associated with the distribution of
intrinsic angular momentum, and the spacelike 1-form field $\mathbf{K}$ is
associated with the direction of the electron spin. For a detailed discussion
concerning such entities, their relationships and physical interpretation, and
generalizations, see, e.g., \cite{cra,lou1,lou2,ro1,holl,hol}.
Lounesto spinor field classification --- representation independent --- is given by the following spinor field
classes \cite{lou1,lou2}, where in the first three classes it is implicit that
$\mathbf{J}$\textbf{, }$\mathbf{K}$\textbf{, }$\mathbf{S}$ $\neq0$:
\begin{enumerate}
\item[(1)] $\sigma\neq0,\;\;\; \chi\neq0$.
\item[(2)] $\sigma\neq0,\;\;\; \chi= 0$.
\item[(3)] $\sigma= 0, \;\;\;\chi\neq0$.
\item[(4)] $\sigma= 0 = \chi, \;\;\;\mathbf{K}\neq0,\;\;\; \mathbf{S}\neq0$.
\item[(5)] $\sigma= 0 = \chi, \;\;\;\mathbf{K}= 0, \;\;\;\mathbf{S}\neq0$.
\item[(6)] $\sigma= 0 = \chi, \;\;\; \mathbf{K}\neq0, \;\;\; \mathbf{S} = 0$.
\end{enumerate}

The current density $\mathbf{J}$ is always non-zero. Classes (1), (2), and (3)
are called \textit{Dirac spinor fields} for spin-1/2
particles, and classes (4), (5), and (6) are called, respectively, \textit{flag-dipole},
\textit{flagpole} and \textit{Weyl spinor fields}. Majorana and ELKO\footnote{
\emph{Eigenspinoren des Ladungskonjugationsoperators} --- Dual-helicity eigenspinors of the charge conjugation operator.}  spinor fields \cite{ro1,alu1,alu2} are
a particular case of a class-(5) spinor field. A great class of applications can be found in \cite{boe1,boe2,alu1,alu2}. It is worthwhile to point out a
peculiar feature of spinor fields of class (4), (5), and (6): although $\mathbf{J}$ is
always non-zero, we have $\mathbf{J}^{2}=-\mathbf{K}^{2}=0$. 
Although the choices given by Eq.(\ref{5}) is restricted to class-(2) DSFs, we can explore 
other choices for values of $\sigma = \bar\psi\psi$ and $\chi = \bar\psi\g_5\psi$, and also investigate the QSL from the point of view 
of classes (1) and (3) spinor fields.

Now, if instead of class-(2) we consider the class-(3) DSF, in which case the spinor field satisfies the normalization conditions
\begin{equation}
\sigma = \bar\psi\psi = 0,\qquad \chi = \bar\psi\g_5\psi = 1,
\end{equation}
 then
the original QSL can be written as 
\begin{equation}\label{166}
\mathcal{L}_\Psi = 
-\Omega_{ab}\w(\vartheta^a\w\vartheta^b) + 
d[D(\bar\psi\vartheta)\w\g_5\psi\vartheta + \bar\psi\vartheta \w\g_5  D(\vartheta\psi)].
\end{equation}\noi
The class-(3) DSF $\psi$ enters the QSL only at the boundary, and consequently it does not appear in the equations of motion. 
Up to the boundary term, therefore, the Lagrangian  is given by the Einstein-Palatini Lagrangian
\begin{equation}
\mt{L}_\Psi = -\Omega_{ab}\w(\vartheta^a\w\vartheta^b).
\end{equation}\noi 

It is immediate to see that, by considering a class-(1) DSF, characterized by the 
conditions $\sigma\neq 0$ and $\chi\neq 0$,
the most general Holst action, given by
\begin{equation}
 \mt{S}^o_\Psi = \bar\psi\psi\,\int\Omega_{ab}\w\star(\vartheta^a\w\vartheta^b) + \bar\psi\g_5\psi\,\int
\Omega_{ab}\w(\vartheta^a\w\vartheta^b),
\end{equation}
follows naturally. In fact, this action comes from the QSL associated with a class-(1) DSF
\begin{equation} \mt{L}_\Psi = -\bar\psi\psi \Omega_{ab}\w\star(\vt^a\w\vt^b) +\bar\psi\g_5\psi\,\Omega_{ab}\w(\vartheta^a\w\vt^b)
+d[D(\bar\psi\vt)\w\g_5\psi\vt + \bar\psi\vt\w\g_5 D(\vt\psi)].
\end{equation}
The ratio $\sigma/\chi$ --- which 
 measures how much Einstein theory departs from a more general
 covariant theory of gravity --- 
is exactly the Immirzi parameter, as pointed out by Chou, Tung, and Yu \cite{imm}.
The expectation value of $\sigma/\chi$ also allows to introduce a renormalization scale upon quantization. 

Whatever representation we use --- Dirac, Weyl or Majorana --- the terms $\sigma = \bar\psi\psi$ and $\chi = \bar\psi\g_5\psi$ are always real. Up to a normalization constant, 
the Immirzi parameter corresponds to the %lembre que o \chi em [6] = i\chi nosso, pois l'a \gamma_5 = i\gamma_{0123}$.
Ashtekar formalism related to the geometrical nature of the new variables \cite{imm,ash1,ash2,ba1,ba2}. In \cite{tung1} the metric was formally introduced
as the symmetric tensor product 
\begin{equation}
g=\bar{\Psi}\otimes_S(1+i\g_5)\Psi\label{455}\end{equation}\noi  where $\Psi$ is given by Eq.(\ref{35}) and $A\otimes_S B := \me(A\ot B + B\ot A)$. There was shown that, when the metric is \emph{real} --- when $g=\bar{\Psi}\otimes_S \Psi$ --- then the Hilbert-Palatini Lagrangian %, given by $(\theta^a\w\theta^b)\w\star
%\Omega_{ab}$
 is obtained \cite{tung1}. Using the property which defines the Hodge star operator, %$\star \Omega_{ab}:=\tilde\Omega_{ab}\g_5$, 
we see that the Hilbert-Palatini action is indeed the Einstein-Hilbert Lagrangian when the curvature $\Omega_{ab}$ is a homogeneous self-dual 2-form field. Since the formalism developed up to now concerns a metric that takes values on the real numbers, it is immediate to extend such formalism in order to obtain, for instance, the self-dual action given in \cite{tung1}, by considering formally also the pure imaginary 
part of the metric given at Eq.(\ref{455}).

%%%%%%%%%%%%%%%%%%%%%%%%%%%%%%%%%%%%%%
\section{Hamiltonian extra terms}
\label{w5}
We now investigate the  Hamiltonian associated with the QSL extra  terms, assuming that $d\psi\neq 0$. 
The energy-momentum can be identified to the Hamiltonian, that can be obtained from the QSL action $S_\Psi$
given a timelike 1-form field $n\in\sec\la^1(T^*M)$ such that $n\cdot dt =1$. As in \cite{tn1} the QSL
4-covariant Hamiltonian 3-form can be written as
\begin{equation}\label{15}
H(n) = 2[D(\bar\psi n^{})\w\g_5 D(\vartheta\psi) + D(\bar\psi \vartheta)\w\g_5 D(n^{}\psi)],
\end{equation}\noi where $n^{} =  n_a \vartheta^a$.
Using a suitable spinor-curvature identity \cite{tung3}, Eq.(\ref{15}) can be written as
\begin{equation}\label{100}
 H(n) = 2\bar\psi\psi\, n^a E_{ab} \star \vartheta^b + 2d[\bar\psi n^{}\w\g_5 D(\vartheta\psi) 
+ D(\bar\psi \vartheta)\w\g_5 n^{}\psi].
\end{equation}\noi where $-2 E^a_{\;\,b}\w\eta^a = \Omega_{ab}\w\eta^{abc}$ is a 3-form field associated with the Einstein tensor
$E = \star^{-1} E_c\ot\vartheta^c$, where $E_c = \Omega_{ab}\w[e_c\lrcorner\star(\vartheta^a\w\vartheta^b)]$.  
Here it is used the definitions $\eta = \vartheta^0\w\vartheta^1\w\vartheta^2\w\vartheta^3$, $\eta_a:=e_a\lrcorner \eta$, 
$\eta_{ab}:= e_a\lrcorner \eta_b$, an $\eta_{abc}:=e_a\lrcorner \eta_{bc}$. 

The choice of a  class-(2) DSFs together with the gauge condition $d\psi=0$ was given in \cite{tn1} 
in order to show the analogy between ADM Hamiltonian, the QSL and the Hamiltonian 3-form field associated 
with the Witten positive energy proof \cite{witten}, where the
 spacetime splitting timelike 1-form field is given by 
$n^\mu = \bar\psi\g^\mu\psi$. 
In this case, the boundary term in Eq.(\ref{100}) reads
\begin{equation}
\omega_{ab}\w[n\lrcorner \star(\vartheta^a\w\vartheta^b)].
\end{equation}
\noi The Hamiltonian in Eq.(\ref{100}) can be also immediately
derived from a suitable spinor-curvature identity \cite{tung3}.
Now we want to investigate the Hamiltonian with the more general condition $d\psi\neq 0$.
Initially we consider class-(1) DSFs to keep all possible terms.
Again, like in the Lagrangian formulation, this condition gives rise to more terms at the boundary in Eq.(\ref{100}).
Indeed, if we expand the boundary term
\begin{equation}
d[\bar\psi n^{}\w\g_5 D(\vartheta\psi) 
+ D(\bar\psi \vartheta)\w\g_5 n^{}\psi],\end{equation}\noi in addition to the well known terms obtained in \cite{tn1}, given by 
\begin{equation}\label{101}
-\bar\psi\g_5\psi n_a\,d\vt^a - \bar\psi\psi \epsilon_{abcd}n^c\omega^{bc}\w\vt^d,
\end{equation}
new terms are obtained:
\begin{equation}
\text{\rm ``extra terms'' = } \bar\psi\g_5 d\psi\,{\rm Tr}(2\vt\w n + 3 n\w\omega\w\vartheta) + \bar\psi\g_5\psi\,{\rm Tr}
(\{dn,d\vt\}-2 dn\w\omega\w\vt).
\end{equation}\noi In the case where $\psi$ is a class-(2) DSF --- 
and in particular $\bar\psi\g_5\psi=0$ --- the second term above equals zero.

Although using the normalization conditions in Eq.(\ref{5})  the boundary terms in Eq.(\ref{101}) are led
exactly to the superpotential associated with M\o ller energy-momentum ``tensor'', this is not true 
when we consider a general class of DSF satisfaying $d\psi\neq 0$.

%%%%%%%%%%%%%%%%%%%%%%
\section{Concluding remarks}
\label{secdis}
QSL makes use of a general auxiliary spin-3/2 field that can be expressed as the tensor
product between an auxiliary spinor field $\psi$ and a Clifford-valued 1-form $\theta$. This auxiliary spinor field $\psi$ was first introduced by Witten as a convenient tool in the proof of the positive-energy theorem of Einstein gravity \cite{witten}. When the QSL is required to yield 
 Einstein-Hilbert, Einstein-Palatini, and Holst actions, it follows naturally that the auxiliary spinor-valued 1-form field composing the QSL must be a DSF.  Any other choice of spinor field leads, up to a boundary term, to a null QSL. In other words, the spinor-valued 1-form field of the QSL must necessarily be constituted by a tensor product between a \emph{Dirac spinor field} and a Clifford algebra-valued 1-form. 

Einstein-Hilbert, Einstein-Palatini, and Holst actions actions correspond respectively to a class-(2), class-(3), and class-(1) DSFs. 
Although the choice $d\psi=0$, and the normalization conditions $\sigma = \bar\psi\psi = 1$
and $\chi=\bar\psi\g_5\psi = 0$ --- corresponding to a class-(2) Dirac spinor field --- gives the best option to prove the
equivalence between the QSL and the Lagrangians associated with general relativity and teleparallel gravity, they are restrictive if we are interested in
more general analyses. Also, classes (2) and (3)
of DSFs can be chosen to give the complete QSL Holst action, each one corresponding respectively
to one of its pieces $\bar\psi\psi\int\Omega_{ab}\w\star(\vt^a\w\vt^b)$  or $\bar\psi\g_5\psi\int\Omega_{ab} \w (\vt^a \w \vt^b)$. Furthermore, class-(1) Dirac spinor field gives
alone the complete Holst action, since in this case $\sigma = \bar\psi\psi \neq 0$ and $\chi = \bar\psi\g_5\psi\neq 0$.

An important point is to observe that the number of parameters in the sets of bilinear covariants is seven for class-(1), and six for classes (2) and (3) DSFs. Then, in the most general case, where the QSL gives rise to the action which is the sum of Einstein-Hilbert and Einstein-Palatini actions, it is not possible, from the covariance of the boundary term in Eq.(14), to infer that the orthonormal frame field and the DSF are tied together. In this case the gauge freedom associated with the DSF $\psi$ cannot be related to the invariance of the boundary term under the Lorentz group, implying that $d\psi\neq 0$. In addition to being too limiting, the condition $d\psi=0$ hides important quantum aspects, such as those related with anomalies. Some of these anomalies arise in a quantum field theory via the boundary term in the Lagrangian \cite{z1,z2,h1,bars,bars1}. 
Even in the related class-(1) DSF, such a condition \emph{can never be imposed} because there are \emph{seven} free parameters, and consequently the condition $d\psi=0$ is forbidden.

\section{Acknowledgment}
The authors are grateful to Prof. R.\ Aldrovandi and to Prof. J. Nester for useful comments. They also thank FAPESP, CNPq and CAPES for financial support. R. da Rocha thanks Prof. W. A. Rodrigues, Jr. for his advices concerning Clifford bundles and spinor fields.

\end{document}